\documentclass[12pt,a4paper,tightenlines,nofootinbib]{revtex4}

\usepackage{amssymb}
\usepackage{amsmath}
\usepackage{epsf}
\usepackage{psfrag}

\newcommand{\beq}{\begin{equation}}
\newcommand{\eeq}{\end{equation}}

\def\rL{\rho_\Lambda}
\def\rD{\rho_D}
\def\nablaslash{\not{\hbox{\kern-3pt $\nabla$}}}

\begin{document}

\author{Jaume Garriga$^{1,2}$, Andrei Linde$^{3}$, and Alexander Vilenkin$^{4}$}
\affiliation{$^1$ Departament de F{\'\i}sica Fonamental,
Universitat de Barcelona, Diagonal 647, 08028 Barcelona, Spain}
\affiliation{$^2$ IFAE, Campus UAB, 08193 Bellaterra (Barcelona),
Spain} \affiliation{$^3$ Department of Physics, Stanford
University, Stanford, CA 94305-4060, USA} \affiliation{$^4$
Institute of Cosmology, Department of Physics and Astronomy, Tufts
University, MA 02155, USA}

\title{Dark energy equation of state and anthropic selection}
\date{\today}

\begin{abstract}

We explore the possibility that the dark energy is due to a
potential of a scalar field and that the magnitude and the slope
of this potential in our part of the universe are largely
determined by anthropic selection effects. We find that, in some
models, the most probable values of the slope are very small,
implying that the dark energy density stays constant to very high
accuracy throughout cosmological evolution. In other models,
however, the most probable values of the slope are such that the
slow roll condition is only marginally satisfied, leading to a
re-collapse of the local universe on a time-scale comparable to the lifetime of the sun. In the latter case, the effective equation of state varies
appreciably with the redshift, leading to a number of testable
predictions.

\end{abstract}

\maketitle

\section{Introduction}

It has long been suggested that both the old fine-tuning problem
of the cosmological constant as well as the puzzle of the time
coincidence may find a natural explanation through anthropic
selection effects, in scenarios where the dark energy density
$\rho_D$ is a random variable
\cite{Davies81,Linde84,Sakharov84,Banks84,Barrow86,Linde87,Weinberg87,AV95,Efstathiou95,MSW,GLV,Bludman,Bousso:2000xa,BanksDine,KL03,GV03}.
This possibility can be easily
realized in the context of inflationary cosmology, where the local
value of $\rho_D$ may be determined by stochastic quantum
processes. These processes may lead to rather different values of
$\rho_D$ in distant regions of the universe, separated by
length-scales much larger than the present Hubble radius.

A simple implementation of this idea is obtained \cite{Linde87,GV00}
by assuming that
the dark energy is due to a scalar field $\phi$ (different from
the inflaton field) with a very flat potential $V(\phi)$, which
has a simple zero at $\phi=\phi_0$ with a nonvanishing slope $s\equiv
|V'(\phi_0)|$:
\begin{equation}
V(\phi) = - s (\phi-\phi_0) + O[(\phi-\phi_0)^2],
\label{linear}
\end{equation}
where we have assumed for definiteness that $V'(\phi_0)<0$.
All that is required is that the slow-roll condition
\begin{equation}
|V'|\lesssim H_0^2 M_P, \label{slowroll}
\end{equation}
is satisfied for values of the potential in the relatively narrow
range
\begin{equation}
|V|\lesssim 10^3 M_P^2 H_0^2 \label{range}.
\end{equation}
Here $H_0$ is the present expansion rate and $M_P$ is the reduced
Planck mass, and we are adopting the convention that any contributions
to the vacuum energy (such as a true cosmological term) are included
in the definition of $V(\phi)$. Larger values of $|V|$ are
uninteresting, since they would severely interfere with structure
formation and with the emergence of suitable 
observers. During
inflation, the value of the scalar field $\phi$ is randomized by
quantum fluctuations, and after inflation it stays almost frozen due
to the flatness of the potential. Thus, the local value of the dark
energy density $\rho_D \approx V(\phi)$ will vary from place to place,
but it will stay almost constant in time. In this situation, the
probability for measuring a particular value of $\rho_D$ is determined
by a combination of inflationary dynamics and anthropic selection
effects. As we shall see in the next Section, this approach to the
cosmological constant problems shows remarkable agreement with
observations, even with the crudest of assumptions.

The purpose of the present paper is to extend this analysis to
scenarios where the slope $s$ of the potential is itself a random
variable. Like $\rD$, the measured value of the slope could be
determined by a combination of inflationary dynamics and anthropic
selection effects. A very large slope would cause a big crunch
much before any observers can develop. If the distribution which
is obtained after inflation favors large values of $s$, then a
value of the slope which {\em marginally} satisfies
(\ref{slowroll}) could be the most probable one to observe
\cite{KL03,GV03,Dimopoulos03}.
Marginal slow-roll entails the
consequence that the effective equation of state depends
appreciably on redshift, $p_D= w_s(z) \rD$, through a function
$w_s$ which contains a single parameter: the value of the slope
$s$ in our region of the universe. Thus, the equation of state
(and its time evolution) may ultimately be determined by the
condition that galaxy formation and the emergence of suitable
observers is marginally allowed before the big crunch happens.
Some observational signatures of models with a marginal slope have
been discussed in \cite{KL03,Dimopoulos03,KKLLS,GPVV}.

In Section II we review the case of variable $\rD$ at fixed $s$.
In Section III we discuss two-field models of dark energy, where
both $\rD$ and $s$ are random variables. Our conclusions are
summarized in Section IV.

\section{Variable $\rD$}

\subsection{Prior distribution}

As mentioned in the Introduction, a theory with variable $\rD$ can
be obtained from a scalar field with a very flat potential, as in
Eq. (\ref{linear}). During inflation, the field $\phi$ undergoes a
random walk of step size $\delta\phi \sim H$ for each time
interval $\delta t\sim H^{-1}$, where $H$ is the expansion rate
during inflation. The steps are taken independently on each
horizon volume, and this leads to spatial variation of $\phi$.
The potential is very flat, and appreciable spatial variation of
$\rho_D$ after thermalization will only occur on scales much
larger than the presently observable universe.

In the limiting case when the potential is absolutely flat, the
rate of expansion of the universe does not depend on the value of
the field $\phi$. Then, because of the Brownian motion of the
field $\phi$ during eternal inflation, the field takes all
possible values with equal probability. In other words, the volume
distribution of the field $\phi$ at thermalization does not depend
on the value of the field and takes the form
\beq
\label{probab}
d{\cal P}_* \propto d\phi.
\eeq

When one takes into account that the potential is not entirely
flat, the situation becomes more complicated. The
probability distribution acquires some $\phi$-dependence, which
may be sensitive to a particular choice of the measure of
probability in an eternally inflating universe. This is a rather
delicate issue, see e.g. \cite{LLM,Vilenkin:1998kr}, but the final
results may not be very sensitive to it because of the extreme
flatness of the potentials suitable for the description of dark
energy.  It has been
argued in \cite{GV00} that for a particular choice of the measure,
and provided that certain generic conditions are satisfied, the
volume distribution of the field $\phi$ at thermalization
preserves the simple form (\ref{probab}) in the narrow range of
anthropic interest (\ref{range}).  We shall return to this issue
in a bit more detail in the next Section, where the case with
several dark energy fields is considered. As we shall see,
additional subtleties arise in that context which require further
discussion. For the rest of this Section, we shall assume that we
are indeed in the situation where the flat distribution
(\ref{probab}) is valid.

From the end of inflation until the present time, the field is
heavily overdamped and remains almost frozen, giving a nearly time
independent contribution to $\rho_D$. Thus, the ``prior"
distribution for the dark energy density is given by
\begin{equation}
d{\cal P}_* \propto {d\rho_D \over |V'(\phi)|}.
\label{priorgeneral}
\end{equation}

Consider, for illustration, the simplest linear potential
\begin{equation}\label{quint}
V(\phi) =\alpha \phi \ .
\end{equation}
If the slope of the potential is
sufficiently small (as in most of the models of dark energy),
\begin{equation} \label{alpha}
\alpha \lesssim
10^{-120} M_p^3,
\end{equation}
then, according to Eq.  (\ref{slowroll}), the field $\phi$ practically does not change during the last
$10^{10}$ years, its kinetic energy is very small. Therefore at
the present stage of the evolution of the universe its total potential energy
$V(\phi)$ acts nearly like a cosmological constant.

Similarly, one may consider a model
\begin{equation}
V(\phi) = {1\over 2} \mu^2 \phi^2 + \rL.
\label{quadratic}
\end{equation}
Here $\rL$ is a true cosmological constant, and $\mu^2 \rL <0$, so
that it is possible to have $|\rD|$ very small even if the constant
$|\rL|$ is very large. Eq. (\ref{slowroll}) leads to the condition
\cite{GV00}
\begin{equation}
|\mu| \lesssim 10^{-120} M_P^3 |\rL|^{-1/2}.
\label{constraint}
\end{equation}
Note that the bounds on $\alpha$ and $\mu$ do not
correspond to a fine tuning, but just to a strong suppression.
Possible mechanisms that could naturally account for such small values
of parameters have been discussed in
\cite{Kallosh:1995hi,Weinberg00,GV00,Donoghue00,DV01,Dimopoulos03}.

The potential vanishes at $\phi_0=-2\rL/\mu^2$, and it can be easily
checked that in the vicinity of this point we have
\beq
|V'(\phi)|= s [1+O(\rD/\rL)],
\label{flat1}
\eeq
where $s$ is the slope at $\phi=\phi_0$ and $\rD\approx V(\phi)$.
Since the true cosmological constant is expected to be large in
absolute value, we have $\rD \ll \rL$, and thus $V'(\phi)\approx {\rm
const}$ in the range of interest, so the potential should be well
approximated by a linear function (\ref{linear}). Substitution into
(\ref{priorgeneral}) yields
\begin{equation}
d{\cal P}_* \propto d\rho_D.
\end{equation}
This means that all values of the dark energy density in the range
(\ref{range}) are equiprobable a priori.

A linear potential as a simple model for dark energy was first
considered in \cite{Linde87}. Later it has been argued that this form
of the potential is generic in the narrow anthropic range
\cite{Weinberg00,GV03,Dimopoulos03}.

It is instructive to compare these models to the more traditional
models of dark energy, with potentials of the type $e^{-cQ}$ or
$Q^{-\beta}$, where $Q$ is the quintessence field
\cite{dark}. Generically, the potential of the quintessence field
contains also a cosmological constant $V_0$, which, {\it a priori},
can be arbitrarily large and can have either sign. Thus, these models do
not solve the cosmological constant problem. They also do not solve
the coincidence problem, unless one fine-tunes the parameters of
the potential.

Since the quintessence potentials become
asymptotically very flat, the scalar field $Q$ also
experiences quantum fluctuations during inflation. Therefore, in these
simple models, one should expect that the typical value of the
quintessence field becomes indefinitely large in the process of
eternal inflation. As a result, these models become completely
indistinguishable from the theory with a simple cosmological constant $V_0$.

One can solve all of these problems, including the cosmological
constant problem, by adding the interaction of the field $Q$ with the
curvature scalar, $\xi RQ^2$, and by multiplying the quintessence
potential by $\phi^n$, where $\phi$ is a massless filed \cite{GV00}. 
In the simplest case $n=1$ the potential of the quintessence models
\cite{dark} in the regime when $Q$ changes very slowly start looking
very similar to our simple linear model (\ref{linear}). Because of the new term $\sim \phi V_0$, one can solve the cosmological constant problem in the same way as in the model with the linear potential \cite{Linde87}. The modified quintessence model will have some features distinguishing it from the model (\ref{linear}), but overall it will be much more complicated.

\subsection{Full distribution}

The distribution ${\cal P}_*(\rD)$ cannot be interpreted directly as
the probability for measuring a particular value of $\rD$. If $|\rD|$
is too large, so that it dominates prior to the galaxy formation epoch
$t_G\sim 10^{10}$ yrs, then it will preclude the very existence of
observers, and hence will never be measured. In order to implement
this selection effect quantitatively, it seems reasonable to
assume that we are typical observers in the ensemble of all
observers in our thermalized region. The probability for measuring
a particular value of $\rD$ can thus be taken to be proportional
to the number of civilizations in the universe which measure that
value of $\rD$, and we have \cite{AV95,GV03}
\begin{equation}
d{\cal P(\rD)}\propto \int dM N_{civ}(\rD,M)n(\rD,M) d{\cal
P}_*(\rD).
\label {posterior}
\end{equation}
Here, $n(\rD,M)dM$ is the number density of galaxies of mass in
the interval $dM$ which will ever form in regions where the dark
energy density takes the value $\rD$, and $N_{civ}$ is the number
of civilizations per galaxy. As a rough approximation we may
assume that the integral is dominated by giant galaxies like the
Milky Way, with mass $M\sim M_G \sim 10^{12}M_\odot$, and that
$N_{civ}$ does not depend significantly on $\rD$. For $M\gtrsim
M_G$, we may take $N_{civ}$ to be proportional to the number of
stars in the galaxy, or to the mass of the galaxy, $N_{civ}(M)
\propto M$. Thus, the probability for measuring a particular value
of $\rho_D$ is proportional to the fraction of matter
$f(M>M_G,\rho_D)$ which clusters in objects larger than $M_G$ in
regions with this value of $\rho_D$. In the Press-Schechter
approximation for determining the fraction of clustered matter in
a $\Lambda$CDM model, and restricting attention to positive values
of $\rho_D$, one finds \cite{MSW,GV03} \beq d{\cal P(\rD)}\propto
f(M>M_G,\rho_D) d\rho_D \propto {\rm erfc} \left[.80
y^{1/3}\right] dy. \label{nG} \eeq Here, we have introduced the
variable $y$ which is linearly related to the dark energy density
$$
y=\left[{F(\Omega_{D0}/\Omega_{m0})\over{\sigma_0(M_G)}}\right]^3
{\rD \over{\rho_{D0}}},
$$
where $\sigma_0(M_G)$ is the present linearized density contrast
on the galactic scale, $\Omega_{m0}+\Omega_{D0}=1$, $\Omega_{D0}$
is the present value of $\Omega_D$ in our local region, and the
function $F(x)$ is given by \beq
F(x)={5\over{6}}\left({{1+x}\over{x}}\right)^{1/2}
\int_0^x{dw\over{w^{1/6}(1+w)^{3/2}}}. \eeq

\begin{figure}[tbh]
\psfrag{od}[][r]{$\Omega_{D0}$} \psfrag{s8}[][l]{$\sigma_8$}
\centering \epsfysize=5.5cm \leavevmode \epsfbox{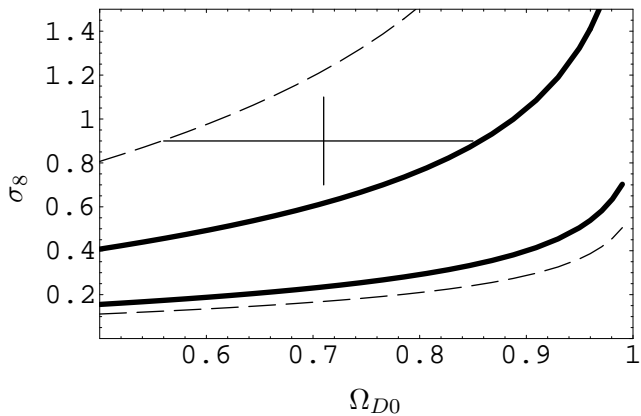}
 \caption{Comparison of anthropic predictions with observations.
 The curves represent the boundaries of the 68\% (solid) and
 95\% (dashed) confidence level regions predicted by the distribution (\ref{nG}).
 The cross represents the values inferred from WMAP observations, with
 $2\sigma$ error bars.} \label{fig1.eps}
\end{figure}

The distribution (\ref{nG}) predicts that $.33<y<6.0$ with 68\%
confidence and $.043<y<16$ at 95\% confidence level. The observed
value of $y$, given by
\begin{equation}
y_0=\left[{F(\Omega_{D0}/\Omega_{m0})\over{\sigma_0(M_G)}}\right]^3
\label{observedvalue}
\end{equation}
is thus expected to fall into these intervals at $1\sigma$ and
$2\sigma$ confidence levels respectively. The boundaries of the
intervals define curves in the $\Omega_{D0}-\sigma_0$ plane. These
curves are shown in Fig. 1, where instead of using the present density
contrast $\sigma_0(M_G)$ on the galactic scale, we use the more
familiar quantity $\sigma_8$. For a given value of $\Omega_{D0}$,
$\sigma_8$ can be obtained from $\sigma_0(M_G)$ if the cosmological
parameters such as the spectral index of density perturbations $n$,
the dimensionless Hubble constant $h$ and the baryon fraction
$\Omega_b$ are known. For these parameters we have taken the central
values given by WMAP, $n=.99, h=.72$ and $\Omega_b=.047$
\cite{wmapestim}. Also shown in the same plot are the values inferred
from WMAP for $\sigma_8$ and $\Omega_{D0}$, which fall well within the
anthropic predictions at the $95\%$ confidence level. Given the
simplicity of the assumptions which have been made, the agreement
between predictions and observations seems quite remarkable.

For negative values of $\rD$, the structure formation proceeds as
usual until the time $t_D\sim (G|\rD|)^{-1/2}$, when the matter
density $\rho_m$ becomes comparable to $|\rD|$. At the moment when
$\rho_m =|\rD|$, the universe stops its expansion and starts
recontraction. The effect of a negative $\rD$ is in many ways similar
to the effect of the slope, and to avoid duplication, we shall not
discuss it here. Interested readers are referred to
Refs.~\cite{KL03,GV03}, where it is argued that the probability for
$\rD<0$ is less or comparable to that of $\rD>0$. In the following
discussion we shall focus on the positive values of $\rD$.

\section{Variable $\rD$ and slope}

In Ref.~\cite{GV03} two of the present authors considered
several predictions of the anthropic approach to the cosmological
constant problems, including the one we have just discussed in the
previous Section. Two other predictions were that the equation of
state should be that of a cosmological constant, $w=-1$, and that the
universe would re-collapse, but not before a trillion years. The
latter predictions were based on the premise that, in a generic
model, the slow-roll condition (\ref{slowroll}) is more likely to
be satisfied by excess, by many orders of magnitude, rather than
marginally. This seems clear in a model such as (\ref{quadratic}),
where the slow-roll parameter is fixed, and a strong suppression
of $\mu$ is required by the constraint (\ref{constraint}). If the
suppression is due to some symmetry, then it is natural to expect
that this symmetry will make the potential flat by excess rather
than marginally. Indeed, the symmetry only knows about
microphysics, and a marginal value $|\mu| \sim 10^{-120} M_P^3
|\rL|^{-1/2}$ would itself represent a coincidence (or tuning)
which requires separate explanation.

However, a marginal value of the slow-roll parameter may be
obtained more naturally in models where the slope is itself a
random variable, by invoking again anthropic selection effects. It
was pointed out in \cite{GV03} that if the prior distribution
favors large $s$, then the most probable values of $s$ could be
the marginal ones.  However, the authors of \cite{GV03} argued (as
we will show, incorrectly) that the prior distribution after
inflation necessarily favors small $s$, and concluded that the
equation of state $w=-1$ should be expected in the general case.
Dimopoulos and Thomas \cite{Dimopoulos03} suggested, on the
contrary, that the prior should generally favor large $s$, but
offered no explicit model to justify this claim. Here, we shall
examine this issue in detail in the context of a specific model.

\subsection{Prior distribution}

A variable slope is easily obtained by considering a model where
we have several fields $\varphi_a$ instead of just one, so that on
the hypersurface $V(\varphi_1,\varphi_2,...,\varphi_n)=0$ the
gradient $s(\varphi) \equiv |\nabla V|$ depends on $\varphi_a$.

During inflation, quantum fluctuations cause a random walk in
field space which covers a distance $|\Delta \varphi_a| \sim H (H
t)^{1/2}$ in time $t$, where $H$ is the inflationary expansion
rate. If we start with some probability distribution $d{\cal
P}_*(\varphi_a, t_i)$ at some initial time $t_i$, the random walk
causes "diffusion" of probability in field space, which tends to
flatten the distribution as inflation proceeds. Hence, if we
neglect the effect of the fields $\varphi_a$ on the expansion of
the universe, we should expect that the volume distribution at the
time of thermalization is given by
\begin{equation}
d{\cal P_*}(\varphi_a)\propto \prod_a d\varphi_a.
\label{simpledist}
\end{equation}
In general, however, the potential of the dark energy fields
$V(\varphi_a)$ does contribute to the expansion, producing a
nontrivial dependence of ${\cal P_*}$ on $\varphi_a$. Even though
$V$ is very small compared with the inflationary energy scale (at
least in the anthropically interesting range), it causes a
"differential expansion" which may accumulate during many
e-foldings, biasing the distribution towards values of the field
where $V(\varphi_a)$ is larger. This is simply because the volume
of regions where the potential is larger grows faster. This leads
to a field dependent distribution,
\begin{equation}
d{\cal P_*}(\varphi_a)=F(\varphi_a) \prod_a d\varphi_a.
\label{calpestar}
\end{equation}
In a region of size $\Delta\varphi$ in the field space, the
characteristic quantum diffusion time is of the order $\tau_q\sim
(\Delta\varphi)^2 /H^3$, and the timescale on which the
differential expansion becomes important is $\tau_{de} \sim
(\Delta H)^{-1} \sim HM_p^2/ \Delta V \sim HM_p^2/|\nabla_\varphi
V| \Delta\varphi$. Diffusion will make the function $F(\varphi_a)$
very smooth (or nearly constant) on scales $\Delta\varphi$ smaller
than the smearing scale
\beq
\Delta\varphi_{smear}\sim H\left({H
M_p^2\over{|\nabla_\varphi V|}} \right)^{1/3},
\label{smears}
\eeq which is obtained by setting $\tau_{de}\sim\tau_q$
\cite{timescales}.  On larger scales, $F$ will generally have a
nontrivial dependence on $\varphi_a$.

In the case we discussed in Section II, where there is a
single dark energy field $\phi$, one can argue that since the
anthropic range for $\rD$ is rather narrow, the corresponding range
of $\phi$ is also limited, and may easily be smaller
than the smearing range (\ref{smears}). This has been used
\cite{GV00,GV03} in order to justify the use of the flat distribution
(\ref{probab}) under certain generic conditions. However, this
justification becomes less clear when we have several dark energy fields,
since the anthropic range does not necessarily correspond to a small
compact region in the field space, and $F$ can vary significantly along
the non-compact directions.

Not much can be said about $F$ in general, since its form depends on
the overall shape of the inflaton and dark energy potential.  (For a
given potential, and with additional assumptions about the measure,
$F$ can in principle be calculated by solving a suitable Fokker-Planck
equation in the formalism of stochastic inflation
\cite{Starob,LLM,Vilenkin:1998kr}). To
simplify our subsequent discussion, here we shall restrict ourselves
to the case where differential expansion is negligible in the field
range of interest. This is achieved for instance through a potential
of the form
\beq
U(\psi,\varphi_a) = U(\psi) + f(\psi) V(\varphi_a),
\eeq
where $\psi$ is the inflaton and $\varphi_a$ are the dark energy
fields. The function $f(\psi)$ is normalized to unity in the
thermalized phase, so that the potential $V(\varphi_a)$ becomes
the dark energy density. If $f(\psi)$ is sufficiently small in the
range of $\psi$ corresponding to most of the inflationary phase,
then the differential expansion can be neglected and the prior
distribution will take the form (\ref{probab}).

The minimal number of fields required to account for variation of
$\rD$ and $s$ is $n=2$.  In this case,
$s=(V_{,1}^2+V_{,2}^2)^{1/2}$ and
\beq
d{\cal P}_*\propto
d\varphi_1 d\varphi_2 \propto |J|^{-1}d\rD ds,
\label{prds}
\eeq
where the Jacobian $J$ is given by
\beq
J\equiv
{\partial(V,s)\over{\partial (\varphi_1,\varphi_2)}}
=s^{-1}\epsilon_{ab} V_{,a} V_{,bc}V_{,c}
\label{J}
\eeq
and $V_{,a}\equiv
\partial V/\partial\varphi_a$.

A simple example is given by
\begin{equation}
V= a \phi + U(\chi),
\label{VU}
\end{equation}
where $\varphi_a=(\phi,\chi)$ and $a$ is a constant.  The properties of the corresponding probability
distribution can be easily understood by expressing it in terms of
the variables $\rD$ and $\chi$. This gives \beq d{\cal P}_*
\propto d\rD d\chi, \eeq where we have used the Jacobian
$\partial(V,\chi)/\partial(\phi,\chi) = \partial V/\partial\phi =
a ={\rm const}$. Thus, for a given interval of $\rD$, all values
of $\chi$ are equally probable. If the range of $\chi$ is
unbounded, then the character of the distribution is determined
largely by the asymptotic behavior of $U(\chi)$. If $U'(\chi)$ is
a growing function of $\chi$, then at large $\chi$ the slope is
$s\approx |U'(\chi)|$, and the probability is dominated by large
values of the slope. For example, for $U(\chi)\propto\chi^2$, we
have $s\propto\chi$, and all values of $s$ are equally probable
(for sufficiently large $s$). Alternatively, if $U'(\chi)\to 0$ at
$|\chi| \to \infty$, then the value $s=a$ is favored.

For a potential of the form
\begin{equation}
V= a \phi + {b\over n}  \chi^n,
\label{V}
\end{equation}
where $a$, $b$ and $n$ are constants, the slope is given by \beq
s=(a^2+ b^2\chi^{2(n-1)})^{1/2}, \label{schi} \eeq and the
Jacobian (\ref{J}) is \beq J\propto \chi^{2n-3}/s \propto
s^{-1}(s^2-a^2)^{{2n-3}\over{2(n-1)}}. \eeq Note that this
expression depends only on $s$, but not on $\rho_D$, and thus the
distribution factorizes:
\begin{equation}
d{\cal P}_*= {\cal P}_*(s) ds d\rD,
 \label{factorized} \eeq with
\beq {\cal P}_*(s)\propto s (s^2-a^2)^{-{{2n-3}\over{2(n-1)}}}.
\label{andrei}
\end{equation}

The slope (\ref{schi}) satisfies $s>a$, so for consistency with
(\ref{slowroll}) we have to require that $a\lesssim H_0^2 M_p$, and
the same logic as above leads us to expect this condition to be
satisfied by excess, that is, $a\ll H_0^2 M_p$.  Then, away from a
small range of $s$ near zero, the distribution (\ref{andrei}) has a
power-law form
\beq
{\cal P}_*(s)\propto s^{-1}s^{1\over{n-1}}.
\eeq

Different behaviors of ${\cal P}_*(s)$ can now be obtained with a
suitable choice of the parameter $n$.  For $n\to\infty$, ${\cal
P}_*(s)\propto s^{-1}$, so all logarithmic intervals of $s$ are
equally probable. (This distribution is also obtained if the power
law $b\chi^n$ is replaced by an exponential function $e^{b\chi}$.)
For $n>1$, the distribution favors large values of $s$, and if $n$
is chosen close to 1, the probability growth towards large $s$ can
be made arbitrarily fast.  With $n>1$, the prior distribution is
non-integrable at large $s$, but in the next subsection we shall
see that an effective cutoff is introduced by the galactic density
factor $n(\rD,M_G)$, so the full distribution (\ref{posterior}) is
integrable.

For readers who are concerned about the appearance of non-integrable
distributions, even at an intermediate stage of the analysis, we note
that the divergence comes from $\chi\to\infty$ and does not occur in
models where the potential has the form (\ref{V}) only in a finite
range of $\chi$. In fact, Eq.~(\ref{V}) with $n=2$ can be thought of
as an expansion of a more general potential (\ref{VU})
in powers of $\chi$ near an extremum of $U(\chi)$.

Finally, for $n<1$, small values of $s$ are favored, and again, by
choosing $n$ close to but below 1, the distribution can be made
arbitrarily steep. In this case, the distribution (\ref{andrei}) is
non-integrable at $s=a$. This singularity is not smoothed out by the
galactic density factor, and the prediction of models with $n<1$ is
that $s=a$ with probability $P=1$. Since $a$ is expected to be very
small, this prediction is observationally indistinguishable from
$s=0$. Once again, the divergence can be cut off if the power-law form of
the potential (\ref{V}) applies only in a finite range of $\chi$.

\subsection{Galactic density}

The probability distribution for measuring given values of $\rD$ and
$s$ is given by a straightforward generalization of
Eq.~(\ref{posterior}),
\begin{equation}
d{\cal P}(\rD,s)\propto \int dM N_{civ}(\rD,s,M)n(\rD,s,M) d{\cal
P}_*(\rD,s).
\label {sposterior}
\end{equation}
If the prior distribution $d{\cal P}_*$ favors small values of $s$,
then we expect the fields $\varphi_a$ to be deep in the slow roll
regime.  In this case, $N_{civ}$ and $n$ are practically independent
of $s$, and we recover the results of Section II.

Suppose now that the prior favors large values of $s$.  To
simplify the discussion, we shall assume that the prior
distribution has a factorized form (\ref{factorized}) with a
power-law distribution for $s$,
\beq
{\cal P}_*(s)\propto s^\beta,
\eeq
$\beta >-1$.  As before, we shall assume that the integral in
(\ref{sposterior}) is dominated by giant galaxies of mass
$M\gtrsim M_G$. Assuming also that for such galaxies
$$
N_{civ}(\rho_D,s,M) = M \nu_{civ}(\rho_D,s),
$$
where $\nu_{civ}$ is the number of civilizations per unit mass, we
have
\beq
d{\cal P}(\rD,s)\equiv {\cal P}(\rD,s)ds d\rD \propto
\nu_{civ}(\rD,s)f(\rD,s,M>M_G) s^\beta ds d\rD.
\label{smposterior}
\eeq
Here, $f$ is the fraction of matter which
clusters in galaxies of size bigger than $M_G$. This fraction
depends on the relative magnitude of three characteristic times:
the galaxy formation timescale $t_G\sim 10^{10}$~yrs, the onset of
vacuum-like dark energy domination $t_D\sim (G\rD)^{-1/2}$, and
the recollapse timescale $t_s$ due to the slope of the potential
(we shall estimate $t_s$ shortly). If $s$ is so small that $t_s$
is the largest of the three times, then the growth of density
fluctuations effectively halts at $t_D$, and the comoving density
of galaxies can be estimated as in Section II. No matter how
small, the slope eventually causes the field to roll down to
negative values of the potential, ending in a big crunch. In the
contracting phase, the density fluctuations start growing again,
and one might think that any galaxies that failed to form at
$t<t_D$ would form then. However, ``galaxies'' that form at this
epoch are likely to be very different from what we call
``galaxies'' now.  At $t_D<t<t_s$, the dark energy density remains
nearly constant, while matter density $\rho_m$ decreases
exponentially with time, so at $t\sim t_s$ it is suppressed by an
exponential factor. Moreover, in the course of the recollapse, the
energy of the scalar field grows much faster than that of matter,
and $\rho_m/\rD$ is further suppressed \cite{negative}. Hence, the
contribution of nonrelativistic matter (like CDM or baryons) to
the mass of bound objects formed during the recollapse is utterly
negligible.

If $t_D$ is the largest of the three times, then the exponential
suppression period is absent, but the rest of the above discussion
still applies, and for $t\gg t_s$ the universe becomes scalar field
dominated. Even at the onset of recollapse, $t\sim t_s$, galaxies as
we know them may not be formed. In our part of the universe, structure
formation effectively stopped at $t\sim t_D$, and the existing
structures evolved more or less in isolation. This may account for the
fact that discs of giant galaxies take their grand-design spiral form
only relatively late, at $z\sim 0.3$. On the other hand, in a
recollapsing universe the clustering hierarchy only speeds up at
$t\gtrsim t_s$, and quiescent discs which may be necessary for the
evolution of fragile creatures like ourselves may never be formed.

This discussion suggests that for $t_D<t_s$, the fraction of matter
clustered in galaxies
can be estimated as
\beq f(\rD,s,M>M_G)\sim f(\rD,M>M_G)
~~~~~~~~~~~~~~~~ (t_D<t_s),
\eeq
where $f(\rD,M>M_G)\equiv
f(r_D,0,M>M_G)$ is the same which we used in the previous Section,
which can be read off from (\ref{nG}). In the opposite case,
$t_D>t_s$, we expect that the density of habitable galaxies does
not exceed the density of galactic-size halos that collapsed prior
to $t_s$. To estimate this density, we shall need an estimate of
$t_s$.

The field equation for $\varphi$ has the form
\beq
{\ddot \varphi}+3({\dot a}/a){\dot \varphi}=s,
\label{phieq}
\eeq
where $a(t)$ is the scale factor. As long as
the dark energy is subdominant, we have $a(t)\propto t^{2/3}$, and
the solution of (\ref{phieq}) is $\varphi=\varphi^{(0)}
+{1\over{6}} s t^2$, where we have imposed the initial condition
${\dot\varphi}\to 0$ at $t\to 0$.  The dark energy density is then
\beq
\rD(t)={1\over{2}}{\dot\varphi}^2-s\varphi={\rD}^{(0)}-{1\over{9}}s^2t^2.
\label{rD1}
\eeq
Assuming first that $t_s<t_D$, we can disregard
$\rD^{(0)}$, and the recollapse begins when the second term in
(\ref{rD1}) becomes comparable to the matter density $\rho_m\sim
1/Gt^2$,
\beq
t_s\sim (M_p/s)^{1/2}.
\label{ts1}
\eeq

Alternatively, if the recollapse occurs after dark energy domination,
then, for $\rD>0$, $a(t)\propto \exp (H_D t)$ with $H_D\sim
(G\rD)^{1/2}$. (Note that for $\rD<0$, the regime $t_s\gg t_D$ does not
exist.) The solution of (\ref{phieq}) is then $\varphi=\varphi^{(0)}
+(s/3H_D)t$, and the dark energy density is
\beq
\rD(t)=\rD^{(0)}-{s^2\over{3H_D}}t.
\label{rD2}
\eeq
This equation applies as long as $\rD$ remains nearly
constant. Recollapse begins when the second term in (\ref{rD2})
becomes comparable to the first,
\beq
t_s\sim {H_D\rD\over{s^2}}\sim {\rD^{3/2}\over{M_p s^2}}.
\label{ts2}
\eeq
The boundary between the two regimes is
\beq
t_s\sim t_D: ~~~~ s\sim\rD/M_p.
\label{sD}
\eeq

Now, it follows from (\ref{ts1}) that the matter density at $t_s$
is $\rho_m(t_s) \sim sM_p$, assuming $t_s\lesssim t_D$. This
suggests that the fraction of matter in habitable galaxies in this regime is
bounded by
\beq
f(\rD,s,M>M_G)\lesssim f(sM_p,M>M_G)
~~~~~~~~~~~~~~~~ (t_D>t_s).
\eeq
In the estimates below, we shall
use the value that saturates this inequality.

We note finally that the dark energy density $\rD$ in
Eqs.~(\ref{sposterior}), (\ref{smposterior}) should be understood as
the value of $\rD$ immediately after inflation, that is, the quantity
denoted by $\rD^{(0)}$ in Eqs.~(\ref{rD1}), (\ref{rD2}).

\subsection{The number of civilizations}

To estimate the dependence of the number of civilizations per unit
mass $\nu_{civ}$ on $\rD$ and $s$, we now have to consider the
role of two other characteristic timescales: $t_*\sim (2-3)\times
10^{10}$~yrs -- the timescale on which most of the main sequence
stars believed to be suitable for life explode as red giants (see
\cite{Livio,GLV} for more discussion and references), and $t_I$ --
the characteristic time needed for intelligent observers to
evolve. $t_I$ is not likely to be much smaller than $t_*$, since
then it is not clear why it took so long for intelligence to
develop on Earth. Carter \cite{Carter} has argued that $t_I\gg
t_*$, since the coincidence $t_I\sim t_*$ is unlikely, considering
that the evolution of life and the evolution of stars are governed
by completely different processes.  Note, however, that some
seemingly unlikely coincidences may occur due to anthropic
selection, $t_G\sim t_D$ being one example. Livio \cite{Livio} has
suggested a simple model illustrating how $t_I\sim t_*$ could
arise. In any case, it seems reasonable to assume that \beq
t_I\gtrsim t_*. \label{tIt*} \eeq The time $t_*$ exceeds $t_G$ by
only a factor $\sim 3$, but it will help to clarify the following
discussion if we proceed as though $t_*\gg t_G$.  This is
justified in part by the fact that we will be comparing densities,
which depend quadratically on time.

Observers can exist only in the time interval \beq t_G<t< {\rm
min}\{t_s,t_*\}, \eeq and since according to (\ref{tIt*}) this
interval is shorter than $t_I$, the number $\nu_{civ}$ is
suppressed by a certain factor. Assuming that the origin of
intelligent life is due to a single and very infrequent random
event which has some constant probability to occur per unit time,
we have \beq \nu_{civ}\propto {{\rm min}\{t_s,t_*\}-t_G}.
\label{Nciv} \eeq  In practice, many steps are necessary for the
development of intelligent life, some of them occurring much more
frequently than others. Assuming that, out of the total number of
steps, there are $k$ of them with typical frequencies smaller than
$1/[{{\rm min}\{t_s,t_*\}-t_G}]$, then Eq. (\ref{Nciv}) should be
modified to
\begin{equation}
\nu_{civ}\propto ({{\rm min}\{t_s,t_*\}-t_G})^k. \label{Ncivn}
\end{equation}
Eq. (\ref{Ncivn}) assumes also that the steps which are needed to
generate intelligence will produce the desired effect regardless
of their time separation. This will not be the situation if there
are relatively frequent catastrophes which occur at intervals
shorter than $({{\rm min}\{t_s,t_*\}-t_G})$, and which are serious
enough to erase memory of any previously achieved steps. In this
case, a linear expression such as (\ref{Nciv}) is more
appropriate.

Our current knowledge about the number of steps $k$ is very poor,
and opinions differ quite vastly. Carter \cite{Carter} has argued
that the effective upper bound to the total time biological
evolution can proceed on earth, $t_{b}$, is likely to be in the
range $t_{b}-t_{e}\lesssim t_e/k$, where $t_e\sim t_*$ is the
actual time intelligent life has taken to develop on earth. This
formula reflects the fact that if many unlikely steps are
necessary, then we are likely to have exhausted most of the
available time before the emergence of intelligent life. If $k$ is
large, Carter's formula seems to indicate that some catastrophe is
awaiting right around the corner which will erase life from earth
within a timescale of order $t_{b}-t_e \ll t_*$. Carter rejected
this possibility, and concluded that $k>2$ was not very likely.
However, Barrow and Tipler \cite{Barrow86} have argued that there
is no reason to reject a relatively imminent doom caused, for
instance, by some instability in the evolution of the earth's
atmosphere. This may render $t_{b}-t_e$ much shorter than $t_*$,
and in this case there is no reason to expect that $k$ should be
small. Cosmic doom of the type we discuss in this paper is yet
another way of obtaining $t_b-t_{e} \ll t_*$, since $t_b<t_s$ and
$t_s$ can in principle be smaller than $t_*$. Qualitatively,
however, our results will not depend strongly on $k$, and for the
rest of the discussion we shall just take $k=1$.

The right-hand side of Eq.~(\ref{Nciv}) takes different forms,
depending on the relative magnitude of $t_s, t_D$ and $t_*$. For
$t_s>t_*$, $\nu_{civ}$ is independent of $s$ and $\rD$, \beq
\nu_{civ}\sim {\rm const}~~~~~~~~~(t_s>t_*). \label{Nciv1} \eeq
For $t_G\ll t_s<t_*$, $\nu_{civ}\propto t_s$, and using
Eqs.~(\ref{ts1}),(\ref{ts2}), we have \beq \nu_{civ}\propto
s^{-1/2}~~~~~~~~~~~(t_s<t_*,t_D), \label{Nciv2} \eeq \beq
\nu_{civ}\propto \rD^{3/2}s^{-2}~~~~~~~~~(t_D<t_s<t_*).
\label{Nciv3} \eeq The boundaries between the different regimes
are given by Eq.~(\ref{sD}) and by \beq t_s\sim
t_*~~~(\rD<1/Gt_*^2): ~~~~ s\sim M_p/t_*^2, \label{s*1} \eeq \beq
t_s\sim t_*~~~(\rD>1/Gt_*^2): ~~~~ s\sim \rD^{3/4}(M_p
t_*)^{-1/2}. \label{s*2} \eeq The corresponding areas in the
$s-\rD$ plane are sketched (not to scale) in Fig.~2.  In our
approximation, the factor $\nu_{civ}$ vanishes for $t_s < t_G$.
(In a more realistic treatment, the density of galaxies would not
strictly vanish for small values of $t_s$. Galaxies would still be
formed at high peaks of the density field, but their number
density would be exponentially suppressed.)   The boundary
$t_s\sim t_G$ is homotetic to the boundary $t_s\sim t_*$ in the
$s-\rho_D$ plane. For $s>\rho_D/M_P$ it simply corresponds to the
vertical line at $s \sim M_P/t_G^2$, and for $s< \rho_D/M_P$ it
follows the curve \begin{equation} s\sim \rho_D^{3/4} (M_P
t_G)^{-1/2}. \label{ht}
\end{equation}
The behavior of $\nu_{civ}$ as a function of $\rD$ and $s$ is
illustrated in Fig. 3.

\begin{figure}[tbh]
\psfrag{fr1}[][]{$M_p^2/t_G^2$} \psfrag{fr2}[][]{$M_P/t_*^2$}
\psfrag{fr3}[][]{$M_P/t_G^2$}\psfrag{tstst}{$t_s\sim t_*$}
\psfrag{tstd}{$t_s\sim t_D$} \psfrag{tstg}{$t_s\sim
t_G$}\psfrag{const}{$\nu_{civ}\propto
const.$}\psfrag{0}{$\nu_{civ}=0$}\psfrag{sm2}{$\nu_{civ}\propto
s^{-2}$} \psfrag{smoh}{$\nu_{civ}\propto s^{-1/2}$}
\psfrag{rD}{$\rho_D$}\psfrag{I}{I} \psfrag{II}{$\ $}
\psfrag{III}{II}\psfrag{IV}{III}\psfrag{V}{IV} \centering
\epsfysize=10cm \leavevmode \epsfbox{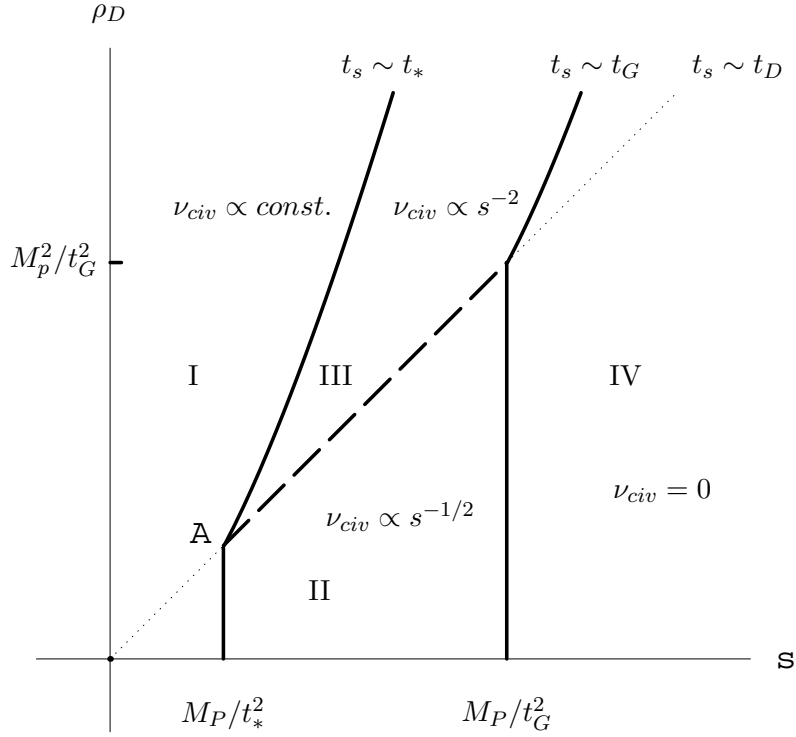}
 \caption{Regions in the $s-\rD$ plane, illustrating the
 different behaviours of $\nu_{civ}$ with $s$.} \label{fig2.eps}
\end{figure}

\begin{figure}[tbh]
\psfrag{Nciv}{$\nu_{civ}$} \psfrag{s}{$s$} \psfrag{rd}{$\rho_D$}
\centering \epsfysize=8cm \leavevmode \epsfbox{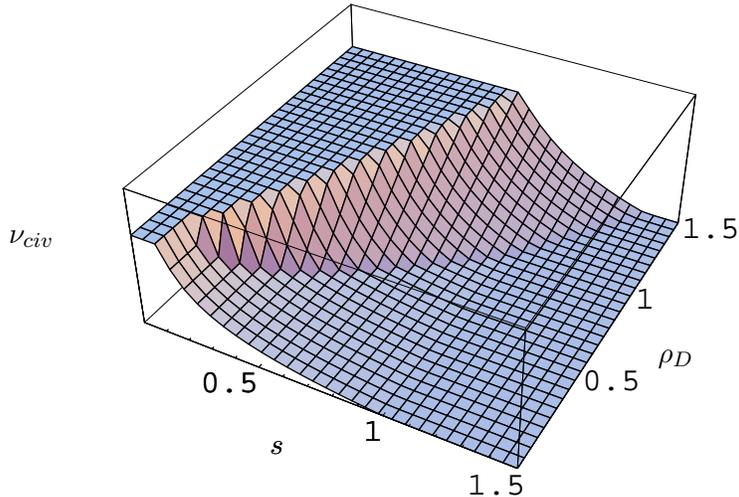}
 \caption{Sketch of $\nu_{civ}$ as a function of $s$ and $\rho_D$.
For definiteness, we have used $t_*= 3 t_G$. The slope $s$ is in
units of $\rho_G/M_p$, whereas $\rho_D$ is in units of $\rho_G$}
 \label{fig3.eps}
\end{figure}

\subsection{Full distribution}

We can now outline the general features of the full distribution
(\ref{smposterior}).  The effect of the galactic density factor
$n(\rD,s,M_G)$ is, roughly, to cut off the distribution outside
the square region $\rD\lesssim 1/Gt_G^2$, $s\lesssim M_p/t_G^2$
(These boundaries correspond to $t_D\sim t_G$ and $t_s\sim t_G$,
respectively.) The fall-off, however, is rather mild, and
$n(\rD,s,M_G)$ extends significantly outside the square. The
cutoff is sharper in the $s$ direction, due to the rapid decline
of $n_{civ}$ with $s$.  In region I ${\cal P}\propto s^\beta$, in
region II ${\cal P}\propto s^{\beta-{1\over{2}}}$, in region III
${\cal P}\propto s^{\beta-2}$, and ${\cal P}\approx 0$ in region
IV.The distribution ${\cal P}$ for $\beta=0$ ($n=2$ in
Eq.~(\ref{V})) is illustrated in Fig. 4. The "logarithmic"
distribution $s \rD {\cal P}(\rD,s)$ is illustrated in Figs. 5 and
6, for $\beta=0$ and $\beta=2$ respectively.

For $-1/2<\beta<1$, the distribution  $s \rD {\cal P}(\rD,s)$ is
peaked along the line separating regions I and III (which
corresponds to $t_s\sim t_*$). As we cross this ``mountain
range'', moving towards larger values of $s$, the probability
function drops in region III. The range terminates at point $A$
where regions I-III meet. For $-1<\beta<-1/2$, the range continues
beyond point $A$ along the boundary between regions I and II.

\begin{figure}[tbh]
\psfrag{P}{${\cal P}$} \psfrag{s}[r][d]{$s$} \psfrag{rd}{$\rho_D$}
\centering \epsfysize=8cm \leavevmode \epsfbox{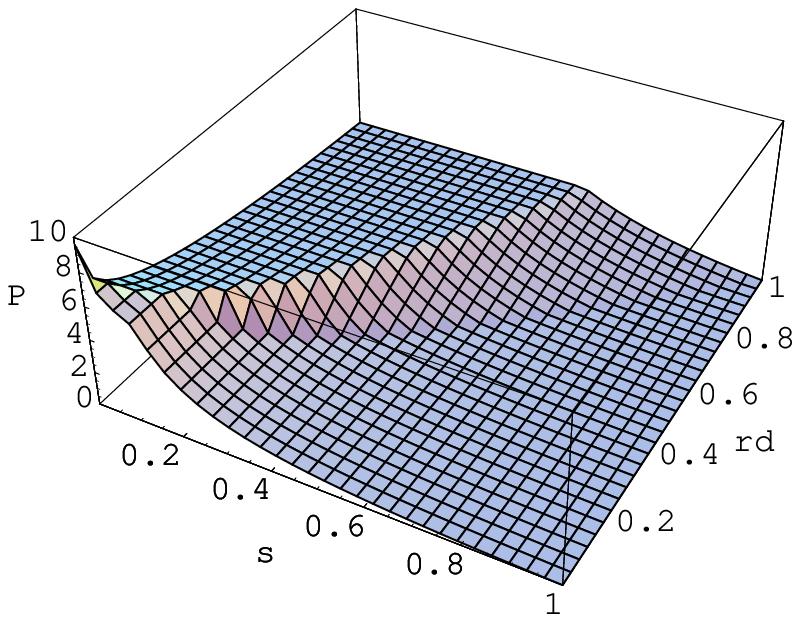}
 \caption{The distribution ${\cal P}(s,\rD)$ as a function of
 $s$ and $\rho_D$, for $\beta=0$. As in Fig. 3, $\rho_D$ is in
 units of $\rho_G$. For this and the following plots, we
have taken $\rho_G\equiv (4/3) M_p^2 t_G^{-2}$, where $t_G=
t_{rec} \sigma_{rec}^{-3/2}$. Here $\sigma_{rec}$ is the density
contrast on the galactic scale at the time of recombination ($t_G$
is then the time it takes for the linearized density contrast to
become equal to 1 in the absence of a dark energy component). With
this choice, the variable $y$ which we used in section II is the
same as $\rD/\rho_G$.}
 \label{fig4.eps}
\end{figure}

\begin{figure}[tbh]
\psfrag{srdP}[r][]{$s \rD {\cal P}$} \psfrag{s}[r][d]{$s$}
\psfrag{rd}{$\rho_D$} \centering \epsfysize=8cm \leavevmode
\epsfbox{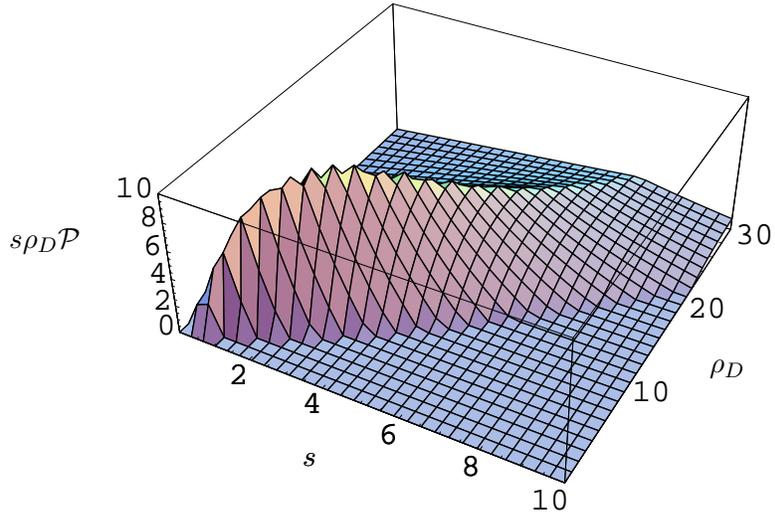}
 \caption{The distribution $s \rD {\cal P}(s,\rD)$ as a function of
 $s$ and $\rho_D$, for $\beta=0$.
 Same conventions as in Fig. 4.}
 \label{fig5.eps}
\end{figure}

\begin{figure}[tbh]
\psfrag{srdP}[r][]{$s \rD {\cal P}$} \psfrag{s}[r][d]{$s$}
\psfrag{rd}{$\rho_D$} \centering \epsfysize=8cm \leavevmode
\epsfbox{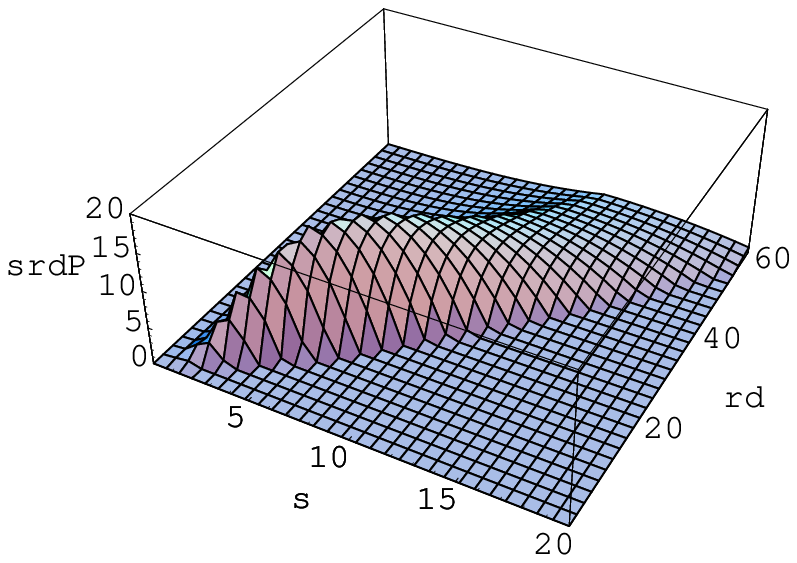}
 \caption{The distribution $s \rD {\cal P}(s,\rD)$ for $\beta=2$.
The rest of parameter values and conventions are as in Figs. 4 and
5.}
 \label{fig5b.eps}
\end{figure}

The probability distribution for $\rD$ can be obtained by
integrating over $s$, \beq {\cal P}(\rD)=\int {\cal P}(\rD,s)ds.
\label{integral}\eeq The character of this distribution depends on
the value of $\beta$.

For $\beta \approx -1$ or smaller, the prior distribution is
peaked on a very narrow strip near $s=0$ in the $s-\rho_D$ plane.
This strip is well within the flat plateau in $\nu_{civ}$ at low
$s$ (see Fig. 3). In this case, integration over $s$ just produces
a constant factor independent of $\rho_D$, and the posterior
distribution coincides with the distribution ${\cal P}(\rD)$ which
is obtained for a true cosmological constant (s=0), given in
(\ref{nG}). For $\beta$ in the range $-1<\beta\lesssim 1$, but not
too close to $-1$, higher values of $s$ come into
play. As shown in Fig. 3, the plateau in $\nu_{civ}$ is broader in
the $s$ direction at high $\rho_D$, and this feature is inherited
by the function ${\cal P}(s,\rD)$ (shown in Fig. 4 for $\beta=0$).
Thus, the integration over $s$ biases the distribution
(\ref{integral}) towards larger $\rD$, relative to the case of a
true cosmological constant. This effect gets stronger as $\beta$
increases.

Finally, for $\beta \gg 1$, the distribution is pushed to the
largest possible values of $s$. For $\beta=2$, the distribution $s
\rD {\cal P}(s,\rD)$ is plotted in Fig. 6. The probability is
concentrated between the curves $t_s\sim t_*$ and $t_s\sim t_G$.
If the value of $\beta$ is further increased, the distribution
gets packed more towards $t_s\sim t_G$, corresponding to
(\ref{ht}). Larger $\beta$ means that larger $s$ is favoured by
the prior distribution, and from (\ref{ht}), this means that the
posterior distribution is peaked at even larger $\rho_D$. The
logarithmic distribution $\rD {\cal P}(\rD)$ is shown in Fig. 7,
for different values of $\beta$, ranging from $-1$ to $2$. The
peak is indeed found to shift to larger values of $\rD$ as we
increase $\beta$.

\begin{figure}[tbh]
\psfrag{rDP}[r][]{$\rD {\cal P}(\rD)$} \psfrag{s}[r][d]{$M_p
s/\rho_G$} \psfrag{rD}{$\rho_D/\rho_G$} \centering \epsfysize=7cm
\leavevmode \epsfbox{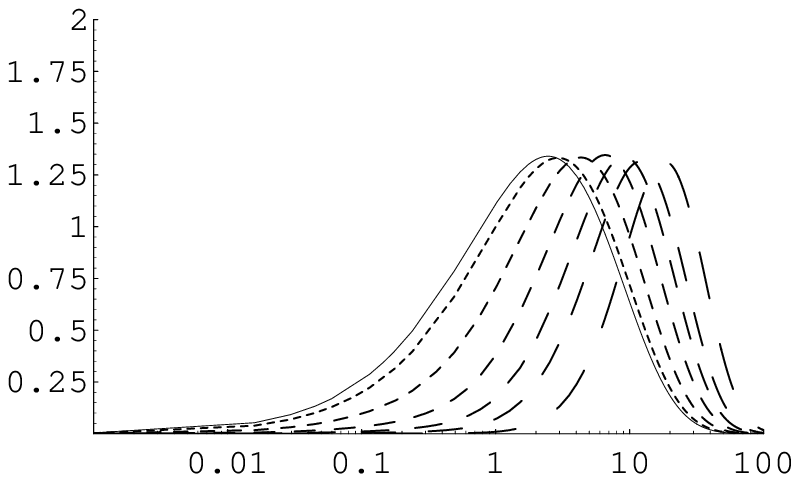}
 \caption{The distribution $\rD {\cal P}(\rD)$ as a function of
$\rho_D/\rho_G$, for different values of $\beta$. The solid line
represents the standard case where $\rD$ is a true cosmological
constant. This is recovered in the case that $\beta=-1$, since in this
case the prior distribution has a non-integrable peak at $s=0$, where
the slope of the potential vanishes. Also plotted are the cases
$\beta=-.9,-.5,0,.5, 1, 2$ (dashed lines, where higher $\beta$
corresponds to longer dashes). The peak of the distribution shifts to
higher $\rho_D$ as $\beta$ is increased.}
\label{fig6.eps}
\end{figure}

Fig. 8 shows the 1-$\sigma$ and 2-$\sigma$ bounds on the variable
$y=\rho_D/\rho_G$ as predicted by (\ref{integral}), as a function
of the parameter $\beta$ in the prior distribution. The central
value inferred from WMAP observations is $y_0 \approx 0.1$, which
for $\beta > -.5$ lies outside the 2-$\sigma$ confidence level
region. However, two things should be noted before jumping to
conclusions. First of all, there is a large uncertainty in the
measured value of $y$ in our region of the universe. For instance,
assuming the WMAP central values for the spectral index $n\approx
.99$, the dimensionless Hubble constant $h\approx .72$ and the
baryon fraction $\Omega_b\approx .047$, and taking into account
the $2-\sigma$ error bars for $\sigma_8$ and $\Omega_D$ (depicted
as a cross in Fig. 1), we find from (\ref{observedvalue}) that the
observed value of $y$ lies in the range $.04< y_0 < .48$. Second,
we should take into account that the predictions represented in
Figs. 1 through 8 refer to the value of the dark energy density
$\rho_D$ at some very early time, when the scalar field is still
frozen by the cosmic expansion. For values of $s \gtrsim
\rho_G/M_P$, the initial value of $\rho_D$ may be larger than the
value at the present time by a sizable fraction. For instance, if
the slope $s$ is such that the kinetic and potential energies of
the dark energy field are approximately equal today (which may be
considered a rather extreme case, although still marginally
consistent with observations \cite{KKLLS}) one finds that the dark
energy density at very early times had to be larger by roughly a
factor of 2. Hence, the initial value of $\rho_D/\rho_G$ in our
region of the universe, may well have been anywhere in the range
$.04 \lesssim y \lesssim 1$. The upper bound in this range is
compatible with a value of $\beta\lesssim 1$ (but not much higher)
at the $2-\sigma$ level.

Given the rough nature of our model and the uncertainties in the
data, our conclusions must be regarded as very preliminary.
Nevertheless, our results do suggest that low values of
$\beta\lesssim -0.5$ should give a better agreement with the data.
This leaves two possibilities: either $\beta<-1$, in which case
the probability is sharply peaked at $s\approx 0$, and all
predictions of Ref.\cite{GV03} for models with a fixed slope
remain in force, or $-1<\beta\lesssim 1$, and then the recollapse
is most likely to occur at $t_s\sim t_*$. Since $t_*$ is
comparable to the present age of the universe, we can already
expect to observe the signs of the slowdown of the cosmic
expansion.

\begin{figure}[tbh]
\psfrag{beta}[r][]{$\beta$} \psfrag{s}[r][d]{$M_p s/\rho_G$}
\psfrag{rD}{$\rho_D/\rho_G$} \centering \epsfysize=7cm \leavevmode
\epsfbox{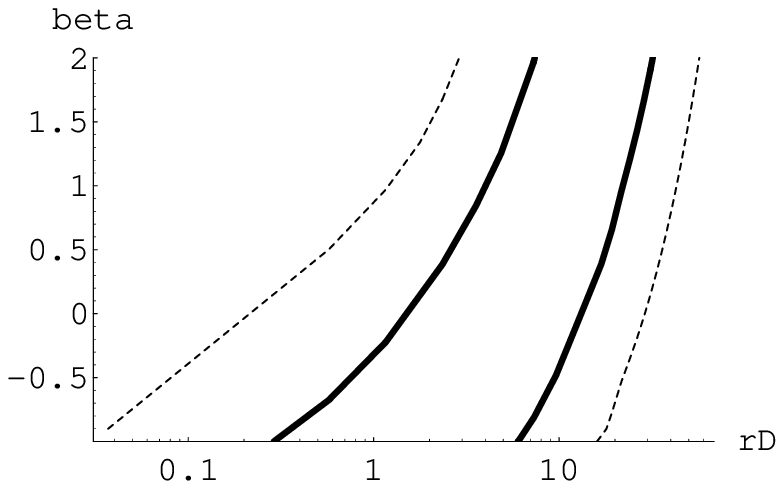}
 \caption{1-$\sigma$ (solid) and 2-$\sigma$ (dashed) bounds on
 the value of the variable $y=\rho_D/\rho_G$, as predicted by (\ref{integral}), as a
 function of the parameter $\beta$ in the prior distribution. Like in the
 previous figures, $\rho_D$ is
 the dark energy density at very early times, and $\rho_G$ is the matter
 density at the time of galaxy formation. The
 central value inferred from WMAP observations is
 $y\approx 0.1$, assuming that the dark energy density
 has remained approximately constant through cosmological evolution
 up to the present time.}
 \label{betabounds.eps}
\end{figure}

 \section{conclusions}

The possibility that the smallness of the observed effective
cosmological constant, as well as the puzzle of the time
coincidence, may be attributed to anthropic selection effects is
rather tantalizing. To implement this idea, one assumes that
$\rho_D$ is a random variable which takes different values in
different parts of the universe (this could be due to stochastic
quantum processes which took place during inflation). Observers
cannot live in regions where $\rho_D$ is too large, since galaxies
cannot form there, so we should not be surprised at the smallness
of the observed $\rho_D$. Also, a simple analysis suggests that
most observers will find themselves in regions which marginally
allow the formation of suitable structures (e.g. galaxies of the
type where observers are most likely to emerge). This would
explain the time coincidence.

A pressing question regarding this scenario is whether it is possible
at all to check its validity. In Ref. \cite{GV03}, two of the
present authors ventured a few generic predictions of the
anthropic approach to the cosmological constant problems. In
particular, following up on the work of Martel, Shapiro and
Weinberg \cite{MSW}, it was argued that the variable $y_0$ defined
in Eq. (\ref{observedvalue}) should be $y_0>.07$ with 95\%
probability. Assuming a COBE normalized scale invariant spectrum
of density perturbations, together with existing estimates for the
baryon density, the anthropic argument suggested that the vacuum
energy density parameter should be somewhat larger than
$\Omega_D=.7$, or that the dimensionless Hubble rate should be
somewhat smaller than $h=.7$. In Section II we have updated the
comparison of predictions with observations by using the
cosmological parameters as obtained from WMAP. Fig. 1 shows
confidence level plots for $\Omega_D$ and $\sigma_8$,
corresponding to the $1-\sigma$ and $2-\sigma$ anthropic
predictions, together with the values inferred from WMAP.

The agreement with current data is rather encouraging, and it will
be interesting to see how it evolves as the level of precision
increases. Note that the confidence level regions in Fig. 1 are
rather broad. This corresponds to a genuine large variance in the
cosmic distribution of $\rho_D$. Hence, one may be led to the
conclusion that future observations will not bring much
excitement, since the overall picture will remain qualitatively
the same even if the observational error bars shrink by a large
factor. Nevertheless, we should recall that a number of
assumptions went into Fig. 1. For instance, we assumed that the
spectral index for scalar fluctuations, the Hubble constant and
the baryon fraction are given by the WMAP central values. We have
also used $w\approx -1$ for the parameter in the dark energy
equation of state. If the values of these parameters turn out to
be different, the curves in Fig. 1 may shift significantly,
putting some pressure on the anthropic explanation.

Also, we must consider the fact that we are quite ignorant about
the conditions which are needed for the emergence of observers, an
obvious drawback of the anthropic approach. However, if we can
encode some of this ignorance in a few unknown physical
parameters, we are in a position where we can predict something
about the values of such parameters. By comparing the theoretical
anthropic predictions with observations, we can find best fit
estimates for the parameters, which may hopefully be confirmed
some day by independent means. We have assumed throughout this
paper that observers emerge predominantly in giant galaxies such
as the Milky Way. This may be a reasonable assumption, but it is
not an established fact by any means. Had we assumed that
observers emerge predominantly in smaller galaxies, which form
earlier on, the agreement with the data would become much worse.
This reasoning was used in Ref. \cite{GV03} to argue that the
conditions for observers to emerge will be found predominantly in
giant galaxies which complete their formation at redshift of order
$z \sim 1$, but not much higher. This prediction seems hard to
check at present, but hopefully much more will be known in the not
so distant future about the properties of galaxies to confirm it
or dispel it.

In this paper we have considered models where the dark energy is
due to the potential energy of several scalar fields. In the case
of a single field, one assumes that the potential has a simple
zero at $\phi=\phi_0$ with nonvanishing slope $s$ (observers
necessarily measure field values close to $\phi_0$, due to
anthropic selection). If the slope is such that the slow-roll
condition (\ref{slowroll}) is satisfied by excess, then the
equation of state will be indistinguishable from that of a true
cosmological constant. But if (\ref{slowroll}) is satisfied only
marginally, then there will be substantial evolution of the
equation of state parameter $w(z)$ with redshift. In models where
the dark energy field has several components, both the dark energy
density $\rho_D$ an the slope $s$ of the potential become random
variables which take different values in distant regions of the
universe (separated by distances larger than the present Hubble
radius). The function $w(z)$ (which is entirely determined by
$\rho_D$ and $s$) will therefore be different in each one of these
regions.

It was argued in Ref. \cite{GV03} that in the case of a single
dark energy field, the slow roll condition (\ref{slowroll}) was
likely to be satisfied by excess, by many orders of magnitude,
rather than marginally. This leads to the predictions that the
equation of state of dark energy is $p_D=-\rho_D$ with very high
accuracy, and that the local universe will re-collapse, but not
before another trillion years. It was also claimed that in generic
two field models one should expect that small slopes would be
favoured by the prior distribution, leading to the same
predictions as in the case of a single field. However, the latter
conclusion was based on an incorrect analysis of the prior
distribution, which we have amended in the present paper.

The prior distribution for the fields at the moment of
thermalization can be obtained in principle from the inflationary
dynamics. In the case of a single field, the anthropically allowed
range (\ref{range}) corresponds to a rather limited region in
field space, and one can argue (under rather mild assumptions)
that the prior distribution of the field will be almost flat
within that range \cite{GV00,Weinberg00}. On the other hand, in the case
where the dark energy potential involves several fields, the range
(\ref{range}) may correspond to a non-compact region in field
space, and the prior distribution may be slowly varying in the
non-compact directions. In this situation, the determination of
the prior distribution from first principles is technically far
more involved (and may require in general some further assumptions
about the choice of the measure in an eternally inflating
universe). Nevertheless, as argued in Section III, a flat
distribution in field space can still be expected provided that
the dark energy potential is sufficiently flat during inflation,
so that its effect on the expansion can be completely neglected.
In this situation, we have shown that there is a class of models
where the prior distribution favors small slopes (in which case
the conclusions of \cite{GV03} hold) but there is an equally broad
class of models where large values of the slope are favoured a
priori.

The measured value of $s$ is restricted by anthropic
considerations, since if it is too large, the local region of the
universe re-collapses before any observers have time to emerge. In
Section III we have attempted to quantify this selection effect,
and we have obtained posterior probability distributions for
$\rho_D$ and $s$. The problem of estimating the abundance of
observers in regions with given values of $\rho_D$ and $s$ has
been split into two parts. In Section III.B we have discussed how
the abundance of suitable galaxies is determined as a function of
$\rD$ and $s$, and in Section III.C, we have analysed how the
number of civilizations may depend on these parameters. There is
of course much room for improvement in these estimates, but even
at the rough level at which they stand, they do illustrate the
fact that a posterior distribution which favors a marginal slope
can easily be obtained in models where the prior favors a large
slope. In this case we find that the universe is likely to turn
around into contraction on a timescale comparable to the lifetime
of the sun. This is a quite exciting prospect since it may lead to
a potentially observable time-dependent equation of state
\cite{KL03,KKLLS,Dimopoulos03,GPVV}.

We finally comment on the string theory motivated picture of a
``discretuum'' of flux compactifications with different values of
$\rD$ \cite{Bousso:2000xa}. Recent work indicates that string theory
does admit vacua with positive $\rD$ \cite{KKLT} and that the
corresponding spectrum of $\rD$ may be rather dense \cite{Douglas},
suggesting the possibility of anthropic selection
\cite{Bousso:2000xa,Susskind,BanksDineLast}. We note, however, that a dense spectrum
of possible values for $\rD$ is only a necessary, but not a sufficient
condition for explaining the value we actually observe. The
probability distribution ${\cal P}(\rD)$ depends on the prior
distribution ${\cal P}_*(\rD)$, and in order to obtain reasonable
agreement with observations, the prior should not be too different
from the flat distribution (\ref{probab}). However, nearby values of
$\rD$ in the discretuum picture correspond to very different values of
the fluxes. The parts of the universe with different values of $\rD$ will have very different evolution histories,
and one might expect that their probabilities will also be rather
different. The arguments we gave in Sections I and IIIA for a
flat prior distribution do not apply to this case. Calculation of probabilities
in the discretuum remains an important problem for
future research.

\

The work
by  A.L. was supported by NSF grant PHY-0244728 and  by the
Templeton Foundation grant No. 938-COS273. The work by J.G. was
supported by CICYT Research Projects FPA2002-3598, FPA2002-00748,
and DURSI 2001-SGR-0061. The work by A.V. was supported by the
National Science Foundation.

\


\begin{thebibliography}{99}

\bibitem{Davies81}
P.~C.~Davies and S.~D.~Unwin,
``Quantum Vacuum Energy And The Masslessness Of The Photon,''
Phys.\ Lett.\ B {\bf 98}, 274 (1981).

%\cite{Linde84}
\bibitem{Linde84}
A.~D.~Linde,
``The Inflationary Universe,''
Rept.\ Prog.\ Phys.\  {\bf 47}, 925 (1984).
%%CITATION = RPPHA,47,925;%%

\bibitem{Sakharov84}
A.~D.~Sakharov,
``Cosmological Transitions With A Change In Metric Signature,''
Sov.\ Phys.\ JETP {\bf 60}, 214 (1984)
[Zh.\ Eksp.\ Teor.\ Fiz.\  {\bf 87}, 375 (1984)].
%%CITATION = SPHJA,60,214;%%


\bibitem{Banks84}
T.~Banks,
``T C P, Quantum Gravity, The Cosmological Constant And All That..,''
Nucl.\ Phys.\ B {\bf 249}, 332 (1985).
%%CITATION = NUPHA,B249,332;%%



\bibitem{Barrow86}
J.D. Barrow and F.J. Tipler, {\it The
Anthropic Cosmological Principle} (Clarendon Press, Oxford, 1986).

\bibitem{Linde87}
A.D. Linde, in {\it 300 Years of Gravitation}, ed. by S.W. Hawking and
W. Israel, Cambridge University Press, Cambridge (1987).

\bibitem{Weinberg87}
S.~Weinberg,
``Anthropic Bound On The Cosmological Constant,''
Phys.\ Rev.\ Lett.\  {\bf 59}, 2607 (1987).

\bibitem{AV95}
A.~Vilenkin,
``Predictions From Quantum Cosmology,''
Phys.\ Rev.\ Lett.\  {\bf 74}, 846 (1995)
[arXiv:gr-qc/9406010].

\bibitem{Efstathiou95}
G. Efstathiou, MNRAS {\bf 274}, L73 (1995).

\bibitem{MSW}
H.~Martel, P.~R.~Shapiro and S.~Weinberg,
``Likely Values of the Cosmological Constant,''
Astrophys.\ J.\  {\bf 492}, 29 (1998)
[arXiv:astro-ph/9701099].

\bibitem{GLV}
J.~Garriga, M.~Livio and A.~Vilenkin,
``The cosmological constant and the time of its dominance,''
Phys.\ Rev.\ D {\bf 61}, 023503 (2000)
[arXiv:astro-ph/9906210].

\bibitem{Bludman}
S.~A.~Bludman,
``Vacuum energy: If not now, then when?,''
Nucl.\ Phys.\ A {\bf 663}, 865 (2000)
[arXiv:astro-ph/9907168].

\bibitem{Bousso:2000xa}
R.~Bousso and J.~Polchinski,
``Quantization of four-form fluxes and dynamical neutralization of the  cosmological constant,''
JHEP {\bf 0006}, 006 (2000)
[arXiv:hep-th/0004134];
J.~L.~Feng, J.~March-Russell, S.~Sethi and F.~Wilczek,
``Saltatory relaxation of the cosmological constant,''
Nucl.\ Phys.\ B {\bf 602}, 307 (2001)
[arXiv:hep-th/0005276].

%\cite{Banks:2003es}
\bibitem{BanksDine}
T.~Banks, M.~Dine and L.~Motl,
``On anthropic solutions of the cosmological constant problem,''
JHEP {\bf 0101}, 031 (2001)
[arXiv:hep-th/0007206].

\bibitem{KL03}
R.~Kallosh, A.~Linde, S.~Prokushkin and M.~Shmakova,
``Supergravity, dark energy and the fate of the universe,''
Phys.\ Rev.\ D {\bf 66}, 123503 (2002)
[arXiv:hep-th/0208156];
R.~Kallosh and A.~Linde,
``M-theory, cosmological constant and anthropic principle,''
Phys.\ Rev.\ D {\bf 67}, 023510 (2003)
[arXiv:hep-th/0208157].


\bibitem{GV03}
J.~Garriga and A.~Vilenkin,
``Testable anthropic predictions for dark energy,''
Phys.\ Rev.\ D {\bf 67}, 043503 (2003)
[arXiv:astro-ph/0210358].

\bibitem{GV00}
J.~Garriga and A.~Vilenkin,
``On likely values of the cosmological constant,''
Phys.\ Rev.\ D {\bf 61}, 083502 (2000)
[arXiv:astro-ph/9908115].






\bibitem{Dimopoulos03}
S.~Dimopoulos and S.~Thomas,
``Discretuum versus continuum dark energy,''
arXiv:hep-th/0307004.




\bibitem{KKLLS}
R.~Kallosh, J.~Kratochvil, A.~Linde, E.~V.~Linder and M.~Shmakova,
``Observational Bounds on Cosmic Doomsday,''
arXiv:astro-ph/0307185.


\bibitem{Kallosh:1995hi}
R.~Kallosh, A.~D.~Linde, D.~A.~Linde and L.~Susskind,
``Gravity and global symmetries,''
Phys.\ Rev.\ D {\bf 52}, 912 (1995)
[arXiv:hep-th/9502069].
%%CITATION = HEP-TH 9502069;%%


\bibitem{GPVV}
J. Garriga, L. Pogosian and T. Vachaspati, in
preparation.

\bibitem{Starob}
A.~Vilenkin,
``The Birth Of Inflationary Universes,''
Phys.\ Rev.\ D {\bf 27}, 2848 (1983).
A.A. Starobinsky, ``Stochastic De Sitter (Inflationary) Stage In The Early Universe,'' in {\it Current topics in Field theory, Quantum
Gravity and Strings}, ed. by H. Vega and N. Sanchez (Springer,
Heidelberg, 1996).

\bibitem{LLM}
A.~D.~Linde, D.~A.~Linde and A.~Mezhlumian,
``From the Big Bang theory to the theory of a stationary universe,''
Phys.\ Rev.\ D {\bf 49}, 1783 (1994);
[arXiv:gr-qc/9306035];
J.~Garcia-Bellido and A.~D.~Linde,
``Stationarity of inflation and predictions of quantum cosmology,''
Phys.\ Rev.\ D {\bf 51}, 429 (1995);
[arXiv:hep-th/9408023];
A.~D.~Linde and A.~Mezhlumian,
``On Regularization Scheme Dependence of Predictions in Inflationary Cosmology,''
Phys.\ Rev.\ D {\bf 53}, 4267 (1996).
[arXiv:gr-qc/9511058].


\bibitem{Vilenkin:1998kr}
A.~Vilenkin,
``Unambiguous probabilities in an eternally inflating universe,''
Phys.\ Rev.\ Lett.\  {\bf 81}, 5501 (1998);
[arXiv:hep-th/9806185];
V. Vanchurin, A. Vilenkin and S. Winitzki, Phys. Rev. {\bf D61},
083507 (2000);
J.~Garriga and A.~Vilenkin,
``A prescription for probabilities in eternal inflation,''
Phys.\ Rev.\ D {\bf 64}, 023507 (2001).
[arXiv:gr-qc/0102090].
%%CITATION = GR-QC 0102090;%

\bibitem{Weinberg00}
S.~Weinberg,
``A priori probability distribution of the cosmological constant,''
Phys.\ Rev.\ D {\bf 61}, 103505 (2000)
[arXiv:astro-ph/0002387].

\bibitem{Donoghue00}
J.~F.~Donoghue,
``Random values of the cosmological constant,''
JHEP {\bf 0008}, 022 (2000)
[arXiv:hep-ph/0006088].

\bibitem{DV01}
G.~R.~Dvali and A.~Vilenkin,
``Field theory models for variable cosmological constant,''
Phys.\ Rev.\ D {\bf 64}, 063509 (2001)
[arXiv:hep-th/0102142].

\bibitem{dark}
 C.~Wetterich,
``Cosmology And The Fate Of Dilatation Symmetry,''
Nucl.\ Phys.\ B {\bf 302}, 668 (1988);
%%CITATION = NUPHA,B302,668;%%
P.~G.~Ferreira and M.~Joyce,
``Cosmology with a Primordial Scaling Field,''
Phys.\ Rev.\ D {\bf 58}, 023503 (1998)
[arXiv:astro-ph/9711102];
B.~Ratra and P.~J.~Peebles,
``Cosmological Consequences Of A Rolling Homogeneous Scalar Field,''
Phys.\ Rev.\ D {\bf 37}, 3406
(1988); I.~Zlatev, L.~M.~Wang and P.~J.~Steinhardt,
``Quintessence, Cosmic Coincidence, and the Cosmological Constant,''
Phys.\ Rev.\ Lett.\  {\bf 82}, 896 (1999)
[arXiv:astro-ph/9807002].
%%CITATION = ASTRO-PH 9807002;%%


\bibitem{wmapestim}
D.~N.~Spergel {\it et al.},
``First Year Wilkinson Microwave Anisotropy Probe (WMAP) Observations: Determination of Cosmological Parameters,''
Astrophys.\ J.\ Suppl.\  {\bf 148}, 175 (2003)
[arXiv:astro-ph/0302209].

\bibitem{timescales}
For a more detailed discussion of the relevant timescales, see
Vanchurin {\it et. al.} in \cite{Vilenkin:1998kr}.

\bibitem{PS}
W.~H.~Press and P.~Schechter,
``Formation Of Galaxies And Clusters Of Galaxies By Selfsimilar Gravitational Condensation,''
Astrophys.\ J.\  {\bf 187} (1974) 425.

\bibitem{negative}
A general analysis of models with negative potentials has been given
in G.~N.~Felder, A.~V.~Frolov, L.~Kofman and A.~V.~Linde,
``Cosmology with negative potentials,''
Phys.\ Rev.\ D {\bf 66}, 023507 (2002)
[arXiv:hep-th/0202017].

\bibitem{Livio}
M. Livio, Ap. J. {\bf 511}, 429 (1999).

\bibitem{Carter}
B. Carter, Phil. Trans. R. Soc. Lond. {\bf A 310}, 347 (1983).

\bibitem{KKLT}
S.~Kachru, R.~Kallosh, A.~Linde and S.~P.~Trivedi,
``De Sitter vacua in string theory,''
Phys.\ Rev.\ D {\bf 68}, 046005 (2003)
[arXiv:hep-th/0301240];
C.~P.~Burgess, R.~Kallosh and F.~Quevedo,
``de Sitter string vacua from supersymmetric D-terms,''
arXiv:hep-th/0309187.
%%CITATION = HEP-TH 0309187;%%



%\cite{Douglas:2003um}
\bibitem{Douglas}
M.~R.~Douglas,
``The statistics of string / M theory vacua,''
JHEP {\bf 0305}, 046 (2003)
[arXiv:hep-th/0303194];
S.~Ashok and M.~R.~Douglas,
``Counting flux vacua,''
arXiv:hep-th/0307049.
%%CITATION = HEP-TH 0307049;%%




%\cite{Susskind:2003kw}
\bibitem{Susskind}
L.~Susskind,
``The anthropic landscape of string theory,''
arXiv:hep-th/0302219.
%%CITATION = HEP-TH 0302219;%%

\bibitem{BanksDineLast} T.~Banks, M.~Dine and E.~Gorbatov,
``Is there a string theory landscape?,''
arXiv:hep-th/0309170.
%%CITATION = HEP-TH 0309170;%%



\end{thebibliography}
\end{document}